\documentclass[12pt, letterpaper]{article}
\usepackage[margin=1in]{geometry}
\usepackage{cite}
\usepackage{amsmath,amssymb,amsfonts}
\usepackage{algorithmic}
\usepackage{graphicx}
\usepackage{textcomp}
\usepackage{multirow}
\usepackage{url}
\usepackage[skip=0.5\baselineskip]{caption}
\usepackage[ruled,vlined]{algorithm2e}
\def\BibTeX{{\rm B\kern-.05em{\sc i\kern-.025em b}\kern-.08em
    T\kern-.1667em\lower.7ex\hbox{E}\kern-.125emX}}
\begin{document}
\LinesNumbered

\title{Auto-tuning of dynamic scheduling applied to 3D reverse time migration on multicore systems}
\author{\'{I}talo A. S. Assis, Jo\~{a}o B. Fernandes, Tiago Barros,\\ Samuel Xavier-de-Souza}

\maketitle

\begin{abstract}
% \textbf{Contextualization}:
Reverse time migration (RTM) is an algorithm widely used in the oil and gas industry to process seismic data. It is a computationally intensive task that suits well in parallel computers.
% \textbf{Gap}:
Methods such as RTM can be parallelized in shared memory systems through scheduling iterations of parallel loops to threads. However, several aspects, such as memory size and hierarchy, number of cores, and input size, make optimal scheduling very challenging.
% \textbf{Purpose}:
In this paper, we introduce a run-time strategy to automatically tune the dynamic scheduling of parallel loops iterations in iterative applications, such as the RTM, in multicore systems. The proposed method aims to reduce the execution time of such applications.
% \textbf{Methodology}:
To find the optimal granularity, we propose a coupled simulated annealing (CSA) based auto-tuning strategy that adjusts the chunk size of work that OpenMP parallel loops assign dynamically to worker threads during the initialization of a 3D RTM application.
% \textbf{Results and discussion}: 
Experiments performed with different computational systems and input sizes show that the proposed method is consistently better than the default OpenMP schedulers, \texttt{static}, \texttt{auto}, and \texttt{guided}, causing the application to be up to 33\% faster. We show that the possible reason for this performance is the reduction of cache misses, mainly level L3, and low overhead, inferior to 2\%. Having shown to be robust and scalable for the 3D RTM, the proposed method could also improve the performance of similar wave-based algorithms, such as full-waveform inversion (FWI) and other iterative applications.

\textbf{keywords:} Auto-tuning, Coupled simulated annealing, Reverse time migration, OpenMP, Dynamic scheduling
\end{abstract}

%\titlepgskip=-15pt

\section{Introduction}
\label{sec:introduction}
    Seismic reflection surveying is the best known and used geophysical method for subsurface imaging. Oil and gas exploration is its main application \cite{Kearey2002}. Its main objective is to generate an image of a subsurface region to identify structures of interest.
    
    Seismic data can go through several processing steps to improve the signal-to-noise ratio (SNR) and the seismic image's resolution. One of the most important of these steps is migration, which is responsible for positioning seismic reflection events in their correct place when imaging the subsurface. In this context, reverse time migration (RTM) \cite{Baysal1983,Kosloff1983} has been widely used as a migration technique to more accurately take into account the wave propagation effects resulting in subsurface images with higher definition.
    
    Simulating wave propagation comprises the majority of an RTM and is computationally intensive, especially for the three-dimensional case. Therefore, the computational cost is the main factor limiting the application of RTM, as well as for several other geophysical algorithms \cite{Zhang2009,Araya-Polo2009}. For this reason, parallel computing techniques have been widely applied to these methods (e.g., \cite{Nunes-do-rosario2015}).
    
    Load balancing is one of the main aspects to be considered in parallel applications. It can be defined as the distribution of the computational load among the available processing resources (e.g., cores, computing nodes). A way to perform load balancing is by dividing the workload in chunks of computation to be distributed among the computational resources either statically or dynamically. In the context of parallel applications, auto-tuning techniques are becoming popular in the sense of obtaining portable near-optimal performance (e.g., when scheduling or load balancing tasks) in parallel programming~\cite{naono2010software}.
    
    Load balancing parallel seismic methods, such as RTM, is especially challenging with the rising interest for heterogeneous machines~\cite{Demetrios2020}. Nevertheless, as we show in this work, even for homogeneous architectures, a static load balancing may not be optimal.
    
    This paper presents an execution time auto-tuning strategy to automatically find an optimal chunk size for OpenMP \cite{OpemMP} dynamic scheduling of parallel loops. This method aims to reduce the run time of iterative applications. For that, we employ a global optimization method called coupled simulated annealing (CSA)~\cite{xavier2010coupled}. This paper also provides numerical experiments where the proposed auto-tuning is applied to a 3D RTM algorithm.
    
    The contribution of this paper to RTM and the scheduling of parallel loops are:
    \begin{enumerate}
        \item the introduction of a novel and robust auto-tuned scheduling strategy for dynamically scheduling parallel loops;
        \item a methodology to perform the parametrization of the proposed method applied to the 3D RTM;
        \item a comparison of the proposed method against the default OpenMP schedulers for the 3D RTM showing that the proposed auto-tuning reduces the number of cache misses;
        \item an overhead analysis showing that the proposed method's overhead was inferior to $2\%$ when employed to the 3D RTM in different computational environments and input sizes;
        \item a consistent reduction in the run-time of the 3D RTM across five different and non-parametrized architectures.
    \end{enumerate}
    
    The outline of the paper is as follows.  We first present the basics of our target application: the RTM (Section \ref{sec:seismic}), the parallelization strategies made available by OpenMP (Section \ref{sec:parallel}) and the optimization method that comprises the proposed auto-tuning, the CSA (Section \ref{sec:csa}). Then, we provide a detailed description of our RTM implementation (Section \ref{sec:rtm}) as well as of the proposed auto-tuning approach (Section \ref{sec:tuning}). Section \ref{sec:results} displays the results of the proposed method in comparison with the standard OpenMP schedules. In Section~\ref{sec:literature} we present a literature review, highlighting the main aspects of our contribution. Section \ref{sec:conclusions} concludes this paper.
    
\section{Reverse Time Migration Formulation}
\label{sec:seismic}
    The seismic reflection method consists of three main steps: acquisition, processing, and interpretation of seismic data. In the acquisition, seismic shots, reflected by subsurface interfaces, are recorded at surface level by receivers. The signal recorded by each detector, from each seismic shot, is called a seismic trace. A set of seismic traces is called a seismogram. Seismograms can be converted to depth estimates of interfaces between different subsurface materials during processing.
    
    After the acquisition step, several techniques can be used to process seismic data. In general, the purpose of processing reflection data is to increase the SNR and improve the vertical resolution of resulting seismic images. Migration is one of the main steps in the seismic data processing. It aims to 1) properly position seismic reflections at the coordinates of the reflector in the subsurface; 2) reduce diffraction effects in the images; 3) improve the spatial resolution.
    
    Modern migration approaches use the seismic wave equation, a partial differential equation describing wave motion, generated by a source in a medium. The scalar equation for 3D acoustic waves is defined as
    \begin{equation} \label{eq:ondaRTM}
    \frac{\partial^2 u(\mathbf{x})}{\partial x_1^2} + \frac{\partial^2 u(\mathbf{x})}{\partial x_2^2} + \frac{\partial^2 u(\mathbf{x})}{\partial x_3^2} = \frac{1}{c(\mathbf{x})^2}\frac{\partial^2 u(\mathbf{x})}{\partial t^2} + s(t),
    \end{equation}
    where $\mathbf{x} = (x_1,x_2,x_3)$ are the spatial dimensions, $u(\mathbf{x})$ is the acoustic pressure, $c(\mathbf{x})$ is the propagation velocity and $s(t)$ is the source function at time $t$.
    
    Spatial and time restrictions should be observed when solving finite differences by a numerical approach \cite{Carcione2002}. These restrictions are defined as:
    \begin{equation} \label{eq:disp}
    \max(\Delta x_1,\Delta x_2,\Delta x_3) \leq \frac{c_{\text{min}}}{W f_{\text{max}}}
    \end{equation}
    and
    \begin{equation} \label{eq:numdins}
    \Delta t \leq \frac{2 \min(\Delta x_1,\Delta x_2,\Delta x_3)}{\pi c_{\text{max}} \sqrt{3}},
    \end{equation}
    where $\Delta x_1$, $\Delta x_2$ and $\Delta x_3$ are the spatial sampling of dimensions $x_1$, $x_2$ and $x_3$, $\Delta t$ is the time sampling; $f_{\text{max}}$ is the maximum frequency of $s(t)$; $c_{\text{min}}$ and $c_{\text{max}}$ are the minimum and the maximum values of $c(\mathbf{x})$; and $W$ is the number of grid points per minimum wavelength. According to \cite{Carcione2002}, $W$ must be equal or greater than 4 for high order finite differences schemes. Non-compliance with~\eqref{eq:disp} and~\eqref{eq:numdins} would result in numerical dispersion and instability.
    
    Another important aspect is that the geological model encoded in $c(\mathbf{x})$ must be restricted to a finite number of points on a mesh, even though the Earth is heterogeneous and continuous. In order to represent real boundaries, it is common to apply artificial edges to the model limits, to absorb the energy reaching the borders \cite{Cerjan1985}.
    
    There are several approaches to migrate seismic data. We use migration by finite differences (or wave equation migration), in which the wave equation is approximated by a finite-difference equation, suitable to be solved by a computer as explained above. One of the main migration methods by finite differences is RTM \cite{Baysal1983,Kosloff1983}. In RTM, source and receiver wavefields are propagated forward and backward in time, respectively. RTM imaging relies on the physical property that those pressure waves must correlate at the reflective interfaces. 
    
    The core of an RTM can be divided into three stages. The first stage is the simulation of the propagation of a wavefield resulting from the excitation of a seismic source. The second stage is the backpropagation of wavefields registered in a seismogram. Finally, the third stage is the imaging condition, which is a correlation between the forward and backward propagated wavefields and produces an image of the subsurface. This process is repeated for all the shots of seismic data available.
    
    The propagation and backpropagation steps use the same velocity model shown in (\ref{eq:ondaRTM}) as $c(\mathbf{x})$. This model specifies the wave velocity for each mesh point and represents the different properties of the materials and boundaries, in the volume being imaged.
    
    In the imaging condition stage, the wavefields generated by the propagation of the source and the backpropagation of the observed data are correlated pointwise, at each time interval, to generate an image. Mathematically, it is defined as
    \begin{equation} \label{eq:correlacao}
    I(\mathbf{x}) = \int\limits_{t=0}^{T} u_\text{i}(\mathbf{x},t) \cdot u_\text{r}(\mathbf{x},t)\text{d}t,
    \end{equation}
    where $I(\mathbf{x})$ is the resulting image, $u_\text{i}(\mathbf{x},t)$ is the wavefield propagated with the source excitation, $u_\text{r}(\mathbf{x},t)$ is the wavefield of backpropagated data and $T$ is the total simulation time. The migration of each seismic shot generates an image. These images are stacked to build the total migrated volume.
    
    Each cycle of a seismic survey ends with the interpretation phase. Since both coverage and resolution are better with 3D data, these surveys lead to improved interpretation compared with 2D surveys and are standard today \cite{yilmaz2001b}.
    
\section{Parallelization strategy}
\label{sec:parallel}
    In this work, we used an RTM algorithm implemented with two degrees of parallelization. The first is the migration of different common-shot (CS) gathers, i.e., seismic data with the same shot coordinates, which is implemented with the message passing interface (MPI)~\cite{MPI1994}, for distributed memory environments. The second is the migration of a single CS gather and is performed in shared memory environments, with OpenMP~\cite{OpemMP}.
    
    The proposed work is applied in the second degree of parallelization, where different loops of the RTM operation of each CS gather are parallelized among the cores of a multicore system, with OpenMP. This parallelization is performed by dividing each loop into loops of smaller sizes, which are computed in the different cores of the multicore system. The size of these smaller loops is usually referred to as the \textit{chunk size}. Our work's primary goal is to balance the computation of the smaller loops by the different cores by choosing the proper chunk size, which is known as workload balancing. The proposed load balancing approach is discussed in more detail in Section~\ref{sec:tuning}.
    
    For the parallelization with OpenMP, the \texttt{parallel for} construction was employed. This construction automatically distributes the workload ($N_{\text{loop}}$) among all threads ($N_{\text{threads}}$), in the loop where it is applied. The workload distribution within the threads can be changed by using the OpenMP clause \texttt{schedule} and variable \textit{chunk size}. The different OpenMP workload distributions used in this work are explained next.

    \textbf{Static}: the load is distributed for each thread in fixed data blocks of roughly $N_{\text{loop}}/N_{\text{threads}}$. It is possible to choose the size of these blocks by changing the \textit{chunk size} variable. 

    \textbf{Dynamic}: similar to the \texttt{static} one, with the main difference that, when a thread finishes to compute the work allocated to it and becomes idle, the system automatically assigns more work to this thread, until all the work finishes.
    
    \textbf{Guided}: similar to the \texttt{dynamic} distribution, in the sense that, when a thread becomes idle, the system also allocates more load to this thread. The difference is that the size of the assigned loop subset starts with the value of $N_{\text{loop}}/N_{\text{threads}}$ and is decreased by the system until it reaches the chosen chunk size value. If the chunk size is omitted, the size of the final loop subset is $1$.

    \textbf{Auto}: is the automatic distribution provided by OpenMP. It delegates all the scheduling decisions to the compiler or run-time system.
    
\section{Coupled simulated annealing}
\label{sec:csa}
    Coupled Simulated Annealing (CSA)~\cite{xavier2010coupled} is a global optimization algorithm, based on the well-known simulated annealing (SA) algorithm~\cite{kirkpatrick1983optimization}. The SA algorithm, also a global optimization method, is inspired by the thermodynamic annealing process, which consists of a heat treatment that alters a given material's physical properties. The SA algorithm is employed in minimization (or maximization) problems, where the goal is to obtain the minimum (or maximum) of a specific cost function, namely the \textit{energy} of the annealing process. This work poses a minimization problem.
    
    Briefly, the SA algorithm is divided into the generation of new solutions and the acceptance of these solutions. Algorithm parameters, known as generation and acceptance \textit{temperatures}, control both these stages.  New possible solutions, also known as probe solutions, are generated by a generation temperature function. If a probe solution yields a smaller value of the cost function, this solution is accepted as the new one with probability one; otherwise, this solution is only accepted as the new one with the probability given by a function of the acceptance temperature.
    
    In its turn, the CSA algorithm consists of a set of parallel SA algorithms, known as SA optimizers. Each SA optimizer generates and evaluates a probing solution, updating its current state. The generation and acceptance temperatures are equal for all the different SA instances. The main differences between CSA and SA are that 1) for accepting solutions with higher cost function values, the CSA considers all current solutions, and 2) the acceptance criterion is based on the current solutions and a \textit{coupling} term between these solutions.
    The coupling approach has shown to be capable of reducing the algorithm's sensitivity to initialization parameters and providing information that might steer the overall optimization process toward the global optimum. 
    
    The CSA algorithm employed in this work is the one described by Gon\c{c}alves-e-Silva et al.~\cite{gonccalves2018parallel}, which is implemented as follows. Let $a_i \in \Theta$ and $b_i \in \Omega$ be, respectively, the current and probe solutions of the $i$-th SA optimizer; with $\Theta$ and $\Omega$ being the set of current and probe solutions, respectively, and $i=1,\,\dots,\,m$, where $m$ is the number of elements in both $\Theta$ and $\Omega$. At the $k$-th iteration of the CSA algorithm, the probe solutions are given by
    \begin{equation}\label{eq:probe_csa}
        b_i = a_i + \epsilon_i T_k^{\text{gen}},
    \end{equation}
    where $T_k^{\text{gen}}$ is the generation temperature and $\epsilon_i$ is a random variable sampled from the Cauchy distribution
    \begin{equation}\label{eq:cauchy}
        g(\epsilon,\,T) = \frac{T}{(\epsilon^2+T^2)^{(D+1)/2}},
    \end{equation}
    where $T=T^{\text{gen}}_k$ and $D$ is the dimension of the problem. The rule for updating $T_k^{\text{gen}}$ is also a free choice of the specific CSA implementation. We followed the guidelines from~\cite{gonccalves2018parallel} and used as
    update  $T_{k+1}^{\text{gen}}=0.99999T_k^{\text{gen}}$, with $T_k^{\text{gen}}$ being updated to $99.999\,\%$ of its previous value.
    
    Each solution, $a_i$ and $b_i$, has an associated energy (or cost) value, $E(a_i)$ and $E(b_i)$. The acceptance probability function is defined as:
    \begin{equation}\label{eq:prob_csa}
        A_{\Theta} = \frac{\exp \left( \frac{E(a_i)-\operatorname{max}(E(a_i))_{a_i \in \Theta}}{T_k^{\text{ac}}}\right)}{\gamma},
    \end{equation}
    where $T_k^{\text{ac}}$ is the acceptance temperature and $\gamma$ is the coupling term, given by:
    \begin{equation}\label{eq:couple_csa}
        \gamma = \sum_{\forall a \in \Theta} \exp \left( \frac{E(a)-\operatorname{max}(E(a_i))_{a_i \in \Theta} }{T_k^{\text{ac}}} \right).
    \end{equation}
    If $E(b_i) > E(a_i)$, $a_i$ assumes the value of $b_i$ only if $A_{\Theta} < r$, where $r$ is a random variable sampled from an uniform distribution in the interval $[0,\,1]$. Otherwise, if $E(b_i) < E(a_i)$, $a_i$ assumes the value of $b_i$ with probability one.
    
    As shown by Xavier-de-Souza et al.~\cite{xavier2010coupled}, 
    the CSA performance is improved if the variance of $A_{\Theta}$ is kept close to its maximum value. This variance might be written as
    \begin{equation}\label{eq:sigma_csa}
        \sigma^2 = \frac{1}{m} \sum_{\forall a \in \Theta} A^2_{\Theta} - \frac{1}{m^2}
    \end{equation}
    and lays in the interval
    \begin{equation}\label{eq:sigma_csa2}
        0 \leq \sigma^2 \leq \frac{m-1}{m^2}.
    \end{equation}
    The controlling of this variance value can be accomplished by using the following rule to update the acceptance temperature:
    
    \begin{equation}\label{eq:tac_csa}
    T^{\text{ac}}_{k+1} =
    \left \{
        \begin{array}{clcc}
        T^{\text{ac}}_{k}(1-\alpha), & \text{if} & \sigma^2 <  \sigma_D^2 \\
        T^{\text{ac}}_{k}(1+\alpha), & \text{if} & \sigma^2 \geq \sigma_D^2 \\
        \end{array}
    \right .,
    \end{equation}
    where $\sigma_D^2$ is the desired variance, which should be kept as close as possible to $\frac{m-1}{m^2}$, and $\alpha$ is the acceptance temperature modification rate, usually a value within the interval $(0,\,0.1]$.
    
    The CSA algorithm is parameterized by setting the initial temperature values $T_0^{\text{gen}}$ and $T_0^{\text{ac}}$; the total number of iterations, $N$; and the number of optimizers, $m$. Setting the number of CSA optimizers, $m$, is heavily discussed in~\cite{gonccalves2018parallel}, where it is shown that, for minimizing several functions, good choices for the number of optimizers lie in the range between $m=4$ and $m=10$. In this work, we chose to use $m=4$ to minimize the auto-tuning overhead without compromising the CSA quality performance. 
    Moreover, as shown in~\cite{xavier2010coupled}, the CSA algorithm is very robust to the initial values of the acceptance temperature, $T_0^{\text{ac}}$, presenting satisfactory results for a large range of $T_0^{\text{ac}}$ values. Regarding the initialization of the generation temperature, $T_0^{\text{gen}}$, and the number of iterations, $N$, section~\ref{subsec:csa_param} provides a guideline for setting these parameters when auto-tuning RTM applications.
    %, where we show that even configuring these parameters is quite a simple task when using the proposed auto-tuning algorithm. 
    
    In the proposed auto-tuning approach, the CSA is employed to minimize the execution time of different loops in the RTM algorithm, parallelized with OpenMP, by properly choosing the optimal chunk size for OpenMP parallel loops. Therefore, the cost function $E(a_i)$ is related to the execution time of an OpenMP \texttt{parallel for} construction and the variable $a_i$ is related to the chunk size, in the \texttt{dynamic} OpenMP distribution.
    
\section{Implementation aspects of RTM}
\label{sec:rtm}
    The RTM program developed to test the proposed auto-tuning is introduced by Assis et al. \cite{Assis2019}. It is implemented in C, using a hybrid parallel approach.
    MPI distributes shots among the nodes of a distributed memory system while OpenMP schedules chunks of the 3D mesh representing the spatial domain to cores of a shared memory system.
    
    The wave propagator of our RTM implementation solves the wave equation by the finite difference method (FDM), using an eighth order in space and second order in time stencil. We used non-reflecting boundary condition to absorb the energy at the boundaries as described in \cite{Cerjan1985}. The absorbing boundary coefficients are computed by
    \begin{align}
    \phi(i) = 
    \begin{cases}
    \pi f_{\text{peak}} \Delta t \left( \dfrac{w_\text{i}}{w_\text{b}} \right) ^ 2, & \text{~on~the~borders,} \\ \label{eq:coef1}
    0, &  \text{~otherwise,}
    \end{cases}\\
    \phi(\mathbf{x}) = \phi(x_1) + \phi(x_2) + \phi(x_3){\rm ,} \label{eq:coef2} \\
    \phi_1(\mathbf{x}) = \frac{1}{1+\phi(\mathbf{x})}{\rm ,} \label{eq:coef3} \\
    \phi_2(\mathbf{x}) = 1-\phi(\mathbf{x}){\rm ,} \label{eq:coef4} 
    \end{align}
    where $f_{\text{peak}}$ is the peak frequency of the source, $w_\text{b}$ is the thickness of the absorbing boundary, in number of grid points, and $w_\text{i}$ ranges from 0 to $w_\text{b}$, indicating the shortest distance from a point $(x_1,x_2,x_3)$ to the border's interior edge. Note that, away from the borders, $\phi_1(\mathbf{x}) = \phi_2(\mathbf{x}) = 1$ and we recover the usual FDM solution.
    
    Using a finite difference second order scheme in time and applying the coefficients of (\ref{eq:coef3}) and (\ref{eq:coef4}) to (\ref{eq:ondaRTM}) leads to
    \begin{align} \label{eq:ondaPML}
    u(\mathbf{x},t+\Delta t) = \phi_1(\mathbf{x})\cdot \Bigg\{ 2u(\mathbf{x},t) - \phi_2(\mathbf{x})\cdot u(\mathbf{x},t-\Delta t) \nonumber\\
    + (c(\mathbf{x})\Delta t)^2\cdot\left[\frac{\partial^2 u(\mathbf{x})}{\partial x_1^2} +\frac{\partial^2 u(\mathbf{x})}{\partial x_2^2} + \frac{\partial^2 u(\mathbf{x})}{\partial x_3^2} - s(t)\right] \Bigg\} {\rm ,}
    \end{align}
    where the source $s(t)$ is modeled as a Ricker wavelet~\cite{wang2015frequencies}.
    
    Our RTM code implements the optimal checkpointing strategy, described in \cite{Symes2007,Griewank2000}, to avoid the use of secondary storage and memory. Details of our RTM implementation are shown in Algorithm \ref{alg:rtm}, which is further discussed in Section \ref{sec:tuning}.

\section{CSA-based auto-tuning}
\label{sec:tuning}
    Katagiri \textit{et al}. \cite{fiber} defines three types of auto-tuning: i) install time, when the estimation procedure is affected by machine environments, ii) before execution invocation, when the estimation procedure is affected by user's knowledge, input parameters or the number of processors, for example, and iii) run time, when the estimation is affected by other parameters generated in run time. In this paper, we propose a run-time auto-tuning for adequately determining the size of parallel loops subsets to be dynamically distributed among OpenMP threads.
    
    The modeling of parallel applications is not trivial, especially if this application runs in different architectures. Variations of aspects such as memory size and hierarchy, number of cores, and input size may have very diverse effects in parallel software performance~\cite{Furtunato2020}.
    Since the relation between the chunk size of the parallel loops and the total execution time of a program is unknown, using a stochastic optimization method to find an optimal chunk size is an alternative. Estimating this relation is particularly challenging for the FDM because of the memory access pattern generated by the stencil computations.
    Using a multidimensional stencil means that the access to memory is non-linear at each wave propagation time step, making it more complex to avoid cache misses. For this reason, the proposed auto-tuning employs CSA to find the chunk size that minimizes the execution time.
    
    For all the tests performed in this work, we adopted the following parameterization rules. For the initial acceptance temperature, we adopted one of the values suggested in~\cite{xavier2010coupled}, $T_0^{\text{ac}} = 0.9$. The number of iterations, $N$, and the number of SA optimizers, $m$, influence the convergence and exploration of the solution variables space. Large values of $N$ and $m$ might result in chunk size estimates closer to the global optimum, with the drawback of greater execution times. For setting the number of optimizers, we chose one of the values tested in~\cite{gonccalves2018parallel}, $m=4$.  The parameters $\sigma^2_D$ and $\alpha$ do not need to be configured, they can be kept fixed, in all simulations, according to \cite{xavier2010coupled}, with $\sigma^2_D = 0.99\left(\frac{m-1}{m^2}\right)$ and $\alpha=0.005$.
    
    For the CSA-based auto-tuning algorithm, the only parameters that needed to be configured were the generation temperature, $T_0^{\text{gen}}$, and the number of iterations, $N$. In section~\ref{subsec:csa_param} we provide insights on how to configure these parameters. For all the tests that we performed, with different data sets, problem sizes, and machines, the CSA algorithm has shown to be quite robust to the initialization of the parameters. We were able to achieve consistent results by using the same parameter values in all the tests.
    
    Algorithm \ref{alg:rtm} presents the pseudo-code of the proposed implementation of the RTM. Regarding the application of CSA for auto-tuning the RTM, four parallel loops of the RTM have been enabled to use the dynamic scheduler with a chunk size that could be defined by the proposed auto-tuning method:
    i) the forward propagation of the source (Line \ref{l:forwardStart});
    ii) the backward propagation of the observed data (Line \ref{l:backwardStart});
    iii) the insertion of the receivers data (Line \ref{l:inputReceiversStart});
    and iv) the image condition (Line \ref{l:imgCondStart}).
    The optimal checkpointing strategy (Line \ref{l:check}) recomputes some of the forward propagation time steps and can also have its chunk size defined by an auto-tuning method.
    
    \begin{algorithm}
    \caption{Reverse Time Migration with auto-tuning. $ns$ is the number of time steps. $t_i$ is the $i$-th time step in the RTM algorithm.}
    \begin{algorithmic}[1]
    \label{alg:rtm}
    \STATE begin time measurement \label{l:itime}
    \STATE distribute shots among nodes using MPI
    \STATE read RTM parameters
    \STATE initialize checkpointing variables
    \STATE compute absorbing boundaries coefficients
    \STATE initialize auto-tuning parameters
    \STATE \#OpenMP parallel section begin
    \FORALL{shots location}
        \STATE read shot seismogram
        \IF {it is the first shot} \label{l:atrtm}
            \STATE \texttt{autotuning()} (See Algorithm \ref{alg:at}) \label{l:atcall}
        \ENDIF
        \FOR {($t_i = 0$ to $ns-1$)} \label{l:loopTime1Start}
            \STATE \#OpenMP parallel loop using the auto-tuned chunk size in a dynamic distribution
            \FORALL {grid points}\label{l:forwardStart}
                \STATE compute the wavefield
            \ENDFOR\label{l:ForwardEnd}
            \STATE add the source wavelet \label{l:InputSource}
            \IF {($t_i$ is a checkpoint)}
                \STATE save Checkpoint
            \ENDIF
        \ENDFOR\label{l:loopTime1End}
        \FOR {($t_i = ns-1$ to $0$)} \label{l:loopTime2Start}
            \STATE \#OpenMP parallel loop using the auto-tuned chunk size in a dynamic distribution \label{l:openmbackward}
            \FORALL {grid points}\label{l:backwardStart}
                \STATE compute the wavefield
            \ENDFOR\label{l:backwardEnd}
            \STATE \#OpenMP parallel loop using static distribution
            \FORALL {receivers location}\label{l:inputReceiversStart}
                \STATE inject observed data samples at time $t_i$
            \ENDFOR\label{l:inputReceirevsEnd}
            \STATE get forward wavefield at $t_i$ from the checkpoints using the auto-tuned chunk size in a dynamic distribution \label{l:check}
            \STATE \#OpenMP parallel loop using static distribution
            \FORALL {main grid points}\label{l:imgCondStart}
                \STATE perform image condition
            \ENDFOR\label{l:imgCondEnd}
        \ENDFOR\label{l:loopTime2End}
    \ENDFOR
    \STATE \#OpenMP parallel section end
    \STATE reduce all nodes migrated sections
    \STATE finish time measurement \label{l:ftime}
    \end{algorithmic} 
    \end{algorithm}
    
    Since the propagation loops are essentially the same, we only apply the proposed auto-tuning for the forward propagation. The chunk size obtained is then used to the forward propagation, the backward propagation, and the checkpointing. On the other hand, the receivers' insertion and image condition loops are not auto-tuned. These loops have a significantly smaller dimension in comparison with the propagation loops. In our tests, these loops together spent less than $2\%$ of the total execution time. Furthermore, they mostly perform linear access to memory, which is ideal for a static distribution. Its overhead may overcome the benefit of auto-tuning the receivers' data insertion and image condition loops. As shown in Algorithm \ref{alg:rtm} (Line \ref{l:atrtm}), the proposed auto-tuning is performed only for the first shot. All the following shots use the same chunk size computed for the first shot.
    
    Algorithm \ref{alg:at} details the implementation of the proposed auto-tuning. The initial set of solutions (chunk sizes) is randomly chosen in the interval $[50,N_{\text{loop}}/N_{\text{threads}}]$. We disregarded small chunk sizes because of the high overhead to schedule them dynamically. Chunk sizes greater than the chunk size of the standard \texttt{static} distribution ($N_{\text{loop}}/N_{\text{threads}}$) are also not taken into consideration because they would lead to the number of blocks to be less or equal than the number of threads and thus, the distribution would be forced to be static.
    
    For each CSA iteration, each optimizer only measures the execution time of the first time step in the forward propagation, using its current chunk size (Lines \ref{l:timeb} and \ref{l:timee}). As shown in \cite{Barros2018}, the run time of the first time step can accurately represent the total propagation execution time. This first-time step is performed twice (Line \ref{l:twice}) and only the elapsed time of the second repetition is registered (Lines \ref{l:secondb} and \ref{l:seconde}) in order to avoid cache population effects. The CSA then uses those time measures as the cost function values and generates the next set of solutions (Line \ref{l:atcsa}).
    
    The overhead in the proposed auto-tuning algorithm is related to the number of iterations used in the CSA global optimization method, which repeats the first time step of the forward wave propagation for each optimization iteration. For instance, in a wave propagation of $1000$ time steps of both forward and backward wave equations and the use of checkpoints with a forward propagation recomputing rate of $2$ to evaluate the image condition ($4000$ times steps in total), with $100$ CSA iterations, the overhead would be $2.5\,\%$ for one shot. Since the auto-tuning is performed only in the forward wave propagation of the first shot, for more shots, the overhead is shared among the time steps of all shots, being significantly reduced in these cases. We tested only instances of a small number of shots, but it is known that 3D seismic acquisitions performed in complex regions can present hundreds of thousands of shots.
    
    \begin{algorithm}
    \caption{Proposed auto-tuning method, function \texttt{autotuning()} of Algorithm \ref{alg:rtm}. $t_i$ is the $i$-th time step in the RTM algorithm.}
    \begin{algorithmic}[1]
    \label{alg:at}
    \STATE $t_i$ = 0
    \FORALL{(auto-tuning iterations)}\label{l:atWhile}
        \FORALL {optimizers}
            \FOR {(i = 1 to 2)} \label{l:twice}
                \IF {(i == 2)} \label{l:secondb}
                    \STATE time measure begin \label{l:timeb}
                \ENDIF
                \STATE \#OpenMP parallel loop using the current chunk size in a dynamic distribution \label{l:openmforward}
                \FORALL {grid points}
                    \STATE compute the wavefield
                \ENDFOR
                \IF {(i == 2)} \label{l:seconde}
                    \STATE time measure end \label{l:timee}
                \ENDIF
            \ENDFOR
        \ENDFOR
        \STATE CSA generates a new solution for each optimizer from the time measures \label{l:atcsa}
    \ENDFOR
    \STATE return the solution with the lowest cost function
    \end{algorithmic} 
    \end{algorithm}

\section{Numerical experiments}
\label{sec:results}
    Our experiments were conducted in five different computational environments, namely:
    \begin{itemize}
        \item Leuven: Single compute node hosting four sixteen-core AMD Opteron(TM) Processor $6376$ at $2.3~\text{GHz}$ and $256~\text{GB}$ RAM. This equipment is located at the \textit{Universidade Federal do Rio Grande do Norte} (UFRN).
        \item NPAD: $68$ compute nodes. Each compute node hosts two CPUs Intel Xeon Sixteen-Core E5-2698v3 at $2.3~\text{GHz}$ and $128~\text{GB}$ RAM DDR4 2133. It is equipped with a $60~\text{TB}$ Lustre parallel distributed file system. This equipment is located at the High-Performance Computing Center at UFRN (NPAD/UFRN).
        \item Yemoja: $856$ compute nodes. Each compute node hosts two $10$-core Intel Xeon E5-2690 Ivy Bridge v2 at $3~\text{GHz}$. $200$ nodes with $256~\text{GB}$ of RAM and $656$ nodes with $128~\text{GB}$ RAM. It is equipped with an $850~\text{TB}$ Lustre parallel distributed file system. This equipment is located at the Manufacturing and Technology Integrated Campus of the National Service of Industrial Training (SENAI CIMATEC).
        \item Ogun: 48 compute nodes. Each compute node hosts two Intel Xeon Gold 6148 at 2.40 GHz and 192 GB RAM DDR4. It is equipped with a 340 TB Lustre parallel distributed file system. This equipment is also located at SENAI CIMATEC.
        \item SDumont: 504 compute nodes. Each compute node hosts two Intel Xeon E5-2695v2 Ivy Bridge at 2.40 GHz and 64 GB RAM DDR4. It is equipped with a 640 TB Lustre parallel distributed file system. This equipment is located at the National Laboratory of Scientific Computing (LNCC).
    \end{itemize}
    
    The compiler used was \textit{gcc}, using its default run-time system (\textit{libgomp}) and the optimization flag \textit{-O3}. We used the most recent version of \textit{gcc} available for each machine, namely $8.1$ for Leuven, $7.3$ for NPAD, $8.2$ for Yemoja, $7.3$ for Ogun and $8.3$ for SDumont. No controls were used to bind threads to processors. Neither the OpenMP \cite{Openmp2015} nor the \textit{libgomp} \cite{Libgomp2019} specifications present how the \texttt{auto} distribution maps the iterations of the parallel loops to the threads. However, it is possible to notice from \textit{libgomp}'s source code \cite{Gitlibgomp2020} that \texttt{auto} distribution maps to the \texttt{static} distribution with a chunk size of roughly $N_{\text{loop}}/N_{\text{threads}}$.
    
    In order to validate the 3D acoustic wave propagator used in our RTM program, we compared a seismic trace computed by our program with the 3D acoustic analytical solution, computed based on~\cite{de1960modification}, in a homogeneous velocity model. The source was a Ricker wavelet with a peak frequency of $20~\text{Hz}$. The distance between source and receiver was $200~\text{m}$. The medium had a constant velocity of $2000~\text{m/s}$. In this experiment, our propagator provided a very accurate approximation to the 3D waveform analytical solution with a mean squared error of $6\times10^{-14}$.
    
    For all the following tests, $f_{\text{peak}} = 20~\text{Hz}$, the time sampling is $1~\text{ms}$, the number of time steps is $3501$, the spatial resolutions are $\Delta x_1 = \Delta x_2 = \Delta x_3 = 10~\text{m}$ and the absorbing border thickness is $50$ points in all directions of the 3D mesh.
    We built $c(\mathbf{x})$ by using a two layers model with a flat interface positioned at the center of the vertical dimension, where the top and bottom layers have velocities of $1400~\text{m/s}$ and $2000~\text{m/s}$, respectively.
    The number of buffers ($n_{\text{b}}$) and checkpoints ($n_{\text{c}}$) depends on the size of the input, as shown in Table \ref{tab:check}. The numbers of buffers were chosen in order to use up to $128~\text{GB}$ of RAM.
    We chose the number of checkpoints optimally according to \cite{Griewank2000}.
    
    \begin{table}
    \centering
    \caption{Number of buffers and checkpoints used in the experiments as function of the input size. The input size does not include the absorbing border. $n_1$, $n_2$ and $n_3$ are the number of samples for the spatial dimensions $x_1$, $x_2$ and $x_3$, being the latter the vertical dimension. $n_2 = n_3 = 401$.}
    \begin{tabular}{|c|c|c|}
    \hline
    \textbf{Input size} $\mathbf{(n_1\times n_2\times n_3)}$ & $\mathbf{n_{\text{b}}}$ & $\mathbf{n_{\text{c}}}$ \\ \hline
    $\mathbf{n_1 = 201}$ & 170 & 3330 \\ \hline
    $\mathbf{n_1 = 401}$ & 100 & 3400 \\ \hline
    $\mathbf{n_1 = 801}$ & 56 & 1848 \\ \hline
    \end{tabular}
    \label{tab:check}
    \end{table}
    
    The execution time measures were taken by using \texttt{MPI\_Wtime()} function. The initial time is measured at the beginning of the algorithm just after launching the processes through MPI (Algorithm \ref{alg:rtm} Line \ref{l:itime}). The final time is measured at the end of the algorithm just before finalizing the processes (Algorithm \ref{alg:rtm} Line \ref{l:ftime}).
    
    \subsection{CSA auto-tuning parameterization}
    \label{subsec:csa_param}
    
    In this section, we explain the methodology employed in CSA parameterization for the auto-tuning of RTM. We provide insights for configuring the CSA parameters in the RTM application. 
    As previously discussed in sections~\ref{sec:csa} and~\ref{sec:tuning}, for the proposed CSA-based auto-tuning of the RTM application, the input parameters that we chose to set were only the generation temperature, $T_0^{\text{gen}}$, and number of iterations, $N$. We fixed the acceptance temperature in $T_0^{\text{ac}} = 0.9$ and the number of optimizers in $m=4$, as suggested in~\cite{xavier2010coupled} and~\cite{gonccalves2018parallel}.
    
    In the following experiment, we show that the configuration of the proposed auto-tuning method is robust to the chosen parameter values and computational architecture, by setting the values of $T_0^{\text{gen}}$ and $N$. In contrast, section~\ref{subsec:performance} shows that the chunk size configuration is quite sensitive to its value and likely to change according to the architecture.
    In that sense, our experiments lead to the conclusion that it may be advantageous to set the auto-tuning parameters once, in a single architecture, instead of performing trial-and-error tests when moving across architectures.
    
    To provide a methodology for choosing the auto-tuning parameters, we propose a parameterization experiment in a specific computational environment. The assumption here is that the parameters obtained in this particular architecture would provide consistent results for other architectures.
    This parameterization experiment yields the values of $T_0^{\text{gen}}$ and $N$ used in all the experiments of auto-tuning the RTM application. Note that a possible extension of the parameterization algorithm experiment is quite straightforward, for the use of the proposed auto-tuning in different applications. The parameterization experiment also shows that the auto-tuning algorithm is reasonably robust to the choice of $T_0^{\text{gen}}$ and $N$ values.
    
    The computational system chosen for the parameterization experiment is composed of two additional compute nodes located in NPAD supercomputer, with the following characteristics:
    \begin{itemize}
        \item One CPU 7250 Intel Xeon Phi with $68$ cores $1.4 \, \text{GHz}$ and $128 \,\text{GB}$ DDR4 ($8 \times 16 \, \text{GB}$).
    \end{itemize}
    
    In this test, the velocity model characteristics were as described in the previous section and its dimensions were $201\times401\times401$. We tested combinations of $N=40, \, 80, \, \text{and} \,160$ and $T_0^{\text{gen}}=1, \, 10, \, 100, \, \text{and} \,1000$.  For each test, we executed $10$ repetitions of the RTM of one seismic shot. We measured the execution time of the entire seismic shot, including the auto-tuning algorithm.
    
\begin{figure}
    	\centering
    	\includegraphics[width=0.99\columnwidth]{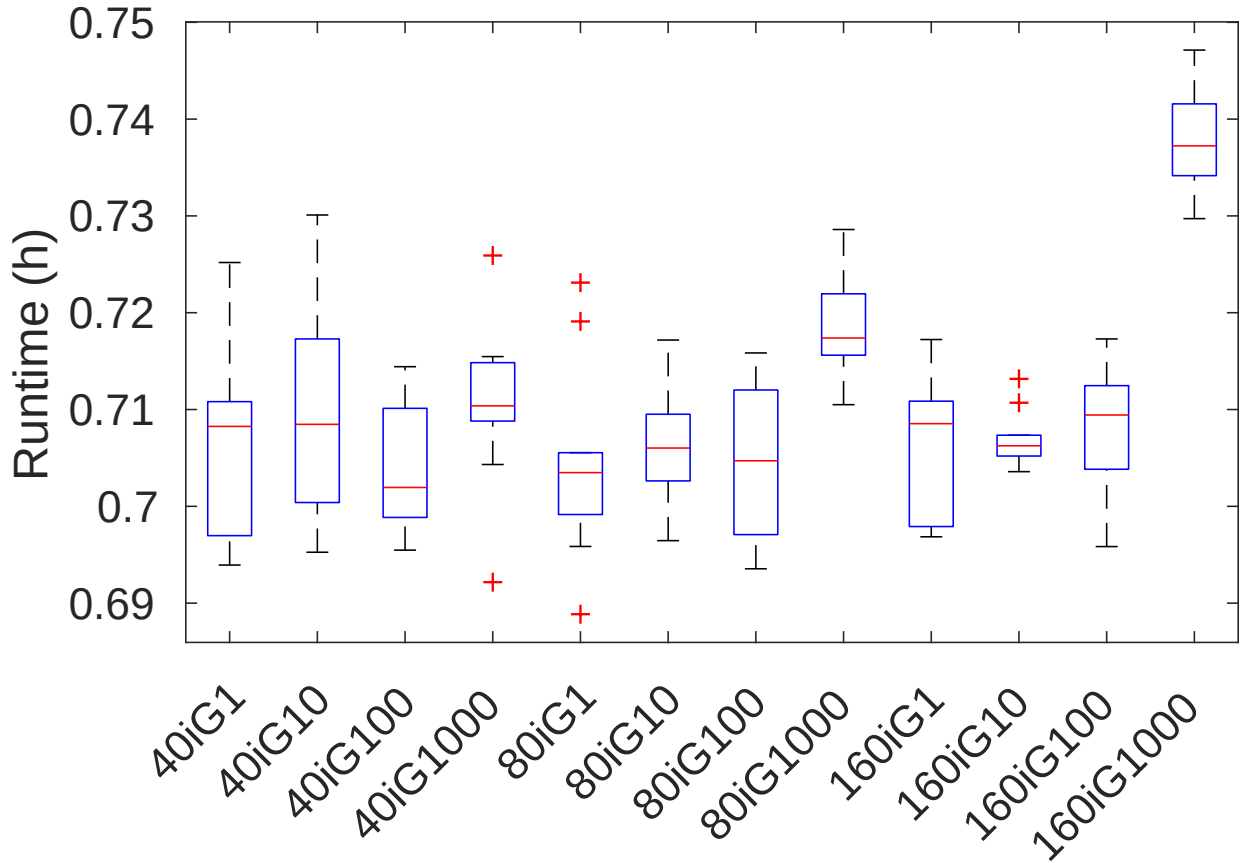}
    	\caption{One-shot RTM run time using the proposed auto-tuning varying the number of CSA iterations ($N=40$, $80$, and $160$), and the initial generation temperature ($T_0^{\text{gen}}=1$, $10$, $100$ and $1000$). The marks on x-axis follows the \textit{XiGY} format meaning $N = X$ and $T_0^{\text{gen}}=Y$. The size of the velocity model was $(n_1\times n_2\times n_3)=201\times401\times401$. These measurements were taken at a CPU 7250 Intel Xeon Phi machine with $68$ cores at $1.4 \, \text{GHz}$.}
    	\label{fig:CSA-param}
	\end{figure}

    Fig.~\ref{fig:CSA-param} shows the obtained results. Most of the shorter execution times occur for the smaller numbers of iterations ($N=40,\, \text{and} \,80$). This fact is most likely because the overhead of the CSA-based auto-tuning is smaller in these cases. It is also possible to observe that in the cases of $N=40,\, \text{and} \,80$ the variance of the executions was quite significant, with the minimum sometimes overlapping among the tests with different parameter combinations. We chose to use $N=40$ as the initial value for the number of iterations since it has a shorter run time. Moreover, for this case, the average run time was quite close for both $T_0^{\text{gen}}=1$, $10$, and $100$. We chose to use $T_0^{\text{gen}}=100$ since it presented the shorter run time among all the executions. 
    
    Note that our parametrization study ran $120$ instances of the RTM, i. e., we have tested $12$ different parameter sets $10$ times each. Also, the number of sets of parameters tested would be the same for different input sizes. Putting it into perspective, the number of RTM executions needed in an exhaustive search for the optimal chunk size depends on the size of the velocity model. For the model size and number of threads used in our experiments, $201\times401\times401$, and $32$, respectively, an exhaustive search would run approximately one million RTM instances.
    
    The CSA configuration parameters used in the remaining experiments are summarized in Table \ref{tab:csaParameters}. We show in Section~\ref{subsec:performance} that the proposed auto-tuning using these parameters was able to outperform the default OpenMP schedulers despite running in different computational architectures and for different input sizes.
    \begin{table}
        \centering
        \caption{CSA parameters used in the numerical experiments.}
        \begin{tabular}{|c|c|c|c|}
            \hline
            $T_0^{\text{gen}}$ & $T_0^{\text{ac}}$ & $N$ & $m$ \\ \hline
            100 & 0.9 & 40 & 4 \\ \hline
        \end{tabular}
        \label{tab:csaParameters}
    \end{table}
    
    \subsection{Performance analysis}
    \label{subsec:performance}
    
    In this subsection, we present three different types of performance experiments. The first type presents the speedup of the 3D RTM when using the proposed method in comparison with OpenMP schedulers \texttt{auto}, \texttt{static}, and \texttt{guided} for different computational architectures and sizes of the velocity model. The second type shows the absolute number of cache misses of the 3D RTM when using the proposed method and the OpenMP schedulers \texttt{auto}, \texttt{static}, and \texttt{guided}. Finally, the third type displays the proposed method's overhead measurements in the 3D RTM for different computational architectures and sizes of the velocity model.
    
    \subsubsection{Speedup analysis}
        Following, we present two sets of experiments regarding the speedups of the proposed auto-tuning method.
        
        The objective of the following set of experiments is to measure the performance of the proposed method in comparison with the OpenMP schedulers, \texttt{auto}, \texttt{static}, and \texttt{guided}, with their default chunk size, i. e., when the chunk size is not explicitly specified. All the following measured run times include the overhead of the proposed method.
        
        The results presented in Table \ref{tab:machineComparison} display the performance of the proposed auto-tuning in comparison with the default OpenMP schedulers in the set of five computational architectures described in Section \ref{sec:results}. For all the experiments, the dimension of the velocity model was $(n_1\times n_2\times n_3)=401\times401\times401$. The proposed method presented a superior performance in all 75 tested scenarios but two, namely, when running the one-shot and the two-shots RTM at Yemoja using the OpenMP \texttt{guided} scheduler. Even in these cases, the performance loss was inferior to $2\%$. As Yemoja has the smallest number of cores per node among the machines tested, the maximum chunk size ($N_{\text{loop}}/N_{\text{threads}}$) is the largest of all scenarios, making the search domain wider for the proposed auto-tuning. On the other hand, the overhead of the OpenMP \texttt{guided} scheduler may be reduced as it manages a smaller number of processing units. The proposed method outperformed the default OpenMP \texttt{static} and \texttt{auto} schedulers in all tested scenarios.
        
        \begin{table*}
        \caption{RTM speedup when using the proposed auto-tuning, compared with the OpenMP schedulers \texttt{auto}, \texttt{static}, and \texttt{guided} in five machines (Leuven, Yemoja, SDumont, NPAD, and Ogun), for $1$, $2$, $4$, $8$ and $16$ seismic shots. For the OpenMP schedulers, the chunk size was not explicitly specified. For these experiments, the size of the velocity model was $(n_1\times n_2\times n_3)=401\times401\times401$. Each point is a median of at least five executions.}
        \setlength{\tabcolsep}{2pt} % Default value: 6pt
        \centering
        \footnotesize
        \hspace*{-.5in}\begin{tabular}{|c|c|c|c|c|c|c|c|c|c|c|c|c|c|c|c|c|}
        \cline{3-17}
        \multicolumn{2}{c|}{\multirow{3}{*}{}} & \multicolumn{15}{c|}{\textbf{Computer environments}}\\ \cline{3-17}
        \multicolumn{2}{c|}{} & \multicolumn{3}{c|}{\textbf{Leuven}} & \multicolumn{3}{c|}{\textbf{Yemoja}} & \multicolumn{3}{c|}{\textbf{SDumont}} & \multicolumn{3}{c|}{\textbf{NPAD}} & \multicolumn{3}{c|}{\textbf{Ogun}} \\ \cline{3-17}
        \multicolumn{2}{c|}{} & \textbf{static} & \textbf{auto} & \textbf{guided} & \textbf{static} & \textbf{auto} & \textbf{guided} & \textbf{static} & \textbf{auto} & \textbf{guided} & \textbf{static} & \textbf{auto} & \textbf{guided} & \textbf{static} & \textbf{auto} & \textbf{guided} \\ \hline
        \multirow{5}{*}{\begin{tabular}{c}\setlength{\tabcolsep}{0pt}\textbf{Number}\\\textbf{of}\\\textbf{shots}\end{tabular}} & \textbf{1} & $5.5\%$ & $7.3\%$ & $9.0\%$ & $8.6\%$ & $10.1\%$ & $-1.3\%$ & $16.5\%$ & $16.4\%$ & $2.1\%$ & $23.1\%$ & $22.7\%$ & $6.1\%$ & $15.3\%$ & $15.3\%$ & $6.5\%$ \\ \cline{2-17}
        & \textbf{2} & $4.5\%$ & $5.1\%$ & $7.1\%$ & $7.2\%$ & $6.1\%$ & $-1.8\%$ & $9.9\%$ & $10.2\%$ & $1.6\%$ & $18.8\%$ & $18.4\%$ & $6.3\%$ & $14.7\%$ & $14.7\%$ & $7.9\%$ \\ \cline{2-17}
        & \textbf{4} & $ 2.6\%$ & $4.0\%$ & $8.9\%$ & $4.4\%$ & $5.6\%$ & $1.2\%$ & $6.2\%$ & $6.9\%$ & $1.4\%$ & $16.0\%$ & $16.3\%$ & $8.4\%$ & $13.1\%$ & $12.9\%$ & $8.9\%$ \\ \cline{2-17}
        & \textbf{8} & $ 2.6\%$ & $2.9\%$ & $6.5\%$ & $10.4\%$ & $8.5\%$ & $3.1\%$ & $4.2\%$ & $4.0\%$ & $0.8\%$ & $16.6\%$ & $14.6\%$ & $9.2\%$ & $13.3\%$ & $13.5\%$ & $9.9\%$ \\ \cline{2-17}
        & \textbf{16} & $ 2.7\%$ & $2.9\%$ & $4.1\%$ & $4.5\%$ & $4.5\%$ & $4.8\%$ & $2.7\%$ & $2.9\%$ & $1.1\%$ & $14.2\%$ & $15.0\%$ & $10.0\%$ & $13.1\%$ & $13.3\%$ & $9.5\%$ \\ \hline
        \end{tabular}
        \label{tab:machineComparison}
        \end{table*}
        
\begin{figure}
    		\centering
    		\includegraphics[width=0.99\columnwidth]{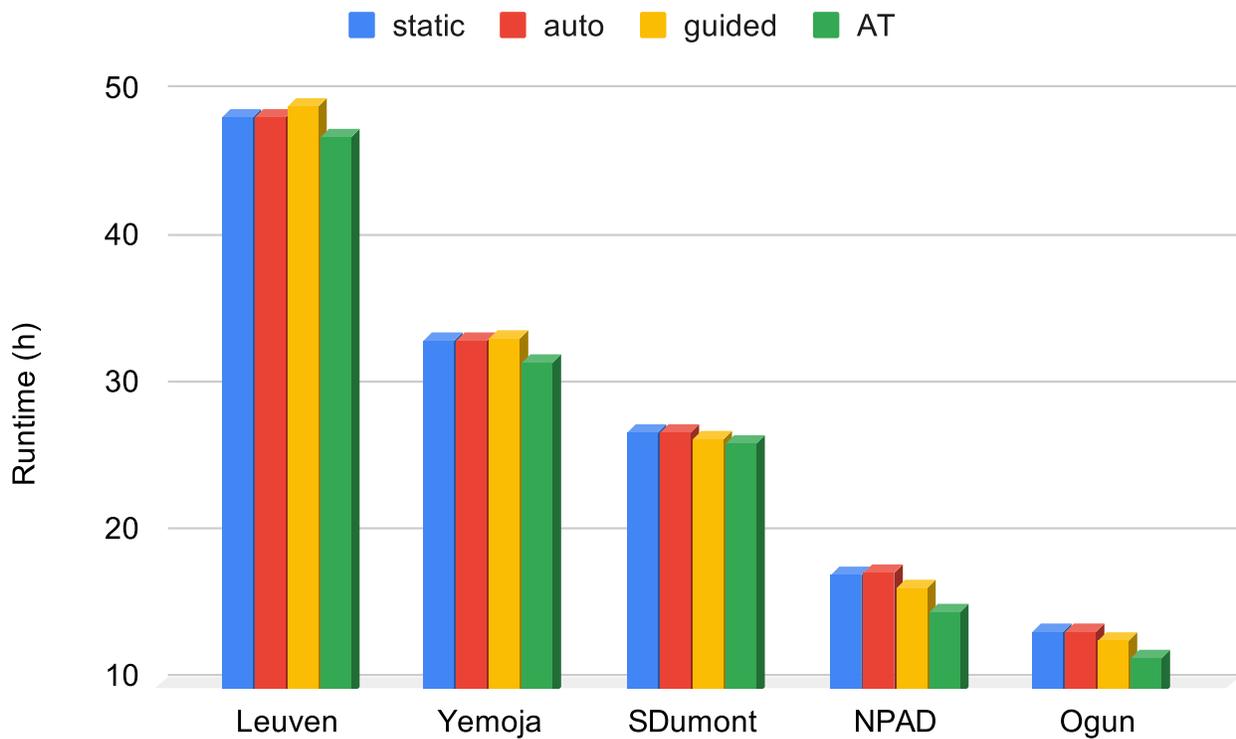}
    		\caption{Execution time of a 16-shot RTM in a single node using the proposed auto-tuning (AT) and the OpenMP schedulers \texttt{auto}, \texttt{static}, and \texttt{guided}, in five machines (Leuven, Yemoja, SDumont, NPAD, and Ogun). For the OpenMP schedulers, the chunk size was not explicitly specified. For these experiments, the size of the velocity model was $(n_1\times n_2\times n_3)=401\times401\times401$. Each measurement presented is a median of at least five executions.}
    		\label{fig:MachineComparison}
		\end{figure}

        Fig. \ref{fig:MachineComparison} details the results shown in Table \ref{tab:machineComparison} for a 16-shot RTM. The proposed method outperforms the three OpenMP schedulers tested regardless of which machine was employed.
        
        The following set of experiments, presented at Table \ref{tab:workloadComparison}, displays the performance of the proposed auto-tuning in comparison with the default OpenMP schedulers, \texttt{static}, \texttt{auto}, and \texttt{guided}, when varying the size of the problem both in means of the dimension of the velocity model and the number of shots. This set of tests was performed at Ogun. The proposed method presents its best performance for larger sizes of the velocity model. Also, for the one-shot RTM, our method's speedup increases with the size of the model.
        
        \begin{table*}
            \caption{RTM speedup when using the proposed auto-tuning, compared with the OpenMP schedulers \texttt{auto}, \texttt{static}, and \texttt{guided} for three input sizes, ($n_1 \times n_2 \times n_3$) equal to $201\times401\times401$, $401\times401\times401$, and $801\times401\times401$, and for $1$, $2$, $4$, $8$ and $16$ seismic shots. For the OpenMP schedulers, the chunk size was not explicitly specified. These tests were performed at Ogun. Each point is a median of at least five executions.}
        	\setlength{\tabcolsep}{4.5pt}
            \centering
            \begin{tabular}{|c|c|c|c|c|c|c|c|c|c|c|}
            \cline{3-11}
            \multicolumn{2}{c|}{\multirow{2}{*}{}} & \multicolumn{3}{c|}{$\mathbf{n_1 = 201}$} & \multicolumn{3}{c|}{$\mathbf{n_1 = 401}$} & \multicolumn{3}{c|}{$\mathbf{n_1 = 801}$} \\ \cline{3-11}
            \multicolumn{2}{c|}{} & \textbf{static} & \textbf{auto} & \multicolumn{1}{c|}{\textbf{guided}} & \textbf{static} & \textbf{auto} & \multicolumn{1}{c|}{\textbf{guided}} & \textbf{static} & \textbf{auto} & \textbf{guided} \\ \hline
            \multirow{5}{*}{\begin{tabular}{c}\setlength{\tabcolsep}{0pt}\textbf{Number}\\\textbf{of}\\\textbf{shots}\end{tabular}} & \textbf{1} & $10.6\%$ & $9.9\%$ & $6.0\%$ & $15.3\%$ & $15.3\%$ & $6.5\%$ & $33.1\%$ & $33.0\%$ & $14.3\%$  \\ \cline{2-11}
            & \textbf{2} & $12.1\%$ & $13.0\%$ & $8.1\%$ & $14.7\%$ & $14.7\%$ & $7.9\%$ & $27.6\%$ & $27.4\%$ & $13.4\%$  \\ \cline{2-11}
            & \textbf{4} & $14.4\%$ & $14.1\%$ & $10.0\%$ & $13.1\%$ & $12.9\%$ & $8.9\%$ & $21.3\%$ & $21.2\%$ & $11.6\%$  \\ \cline{2-11}
            & \textbf{8} & $14.1\%$ & $14.2\%$ & $9.2\%$ & $13.3\%$ & $13.5\%$ & $9.9\%$ & $19.5\%$ & $20.2\%$ & $11.8\%$  \\ \cline{2-11}
            & \textbf{16} & $15.1\%$ & $14.9\%$ & $10.8\%$ & $13.1\%$ & $13.3\%$ & $9.5\%$ & $16.9\%$ & $17.2\%$ & $11.9\%$ \\ \hline
            \end{tabular}
            \label{tab:workloadComparison}
        \end{table*}
        
        The larger the input size, the bigger the chunks of the \texttt{static} scheduler and the initial chunk of the \texttt{guided} scheduler.
        For instance, when running over Ogun, the chunk sizes of the \texttt{static} distribution were approximately $1.9$, $3.1$, and $5.7$ millions of loop iterations for the input sizes of $201\times401\times401$, $401\times401\times401$, and $801\times401\times401$ respectively. By working with larger chunks, the data locality of the static distribution decreases, which explains its performance decrease for larger input sizes. On the other hand, the median of the chunk sizes chosen by the proposed auto-tuning method was $30.3$, $122$, and $232.5$ thousands of loop iterations for the same input sizes, respectively. By processing suitably smaller chunks and distributing them dynamically, the likelihood that all threads work in a contiguous memory range increases, and so does the data locality in the proposed method. This temporal data locality increases the reuse of the data in cache memory.
        
\begin{figure}
    		\centering
    		\includegraphics[width=0.99\columnwidth]{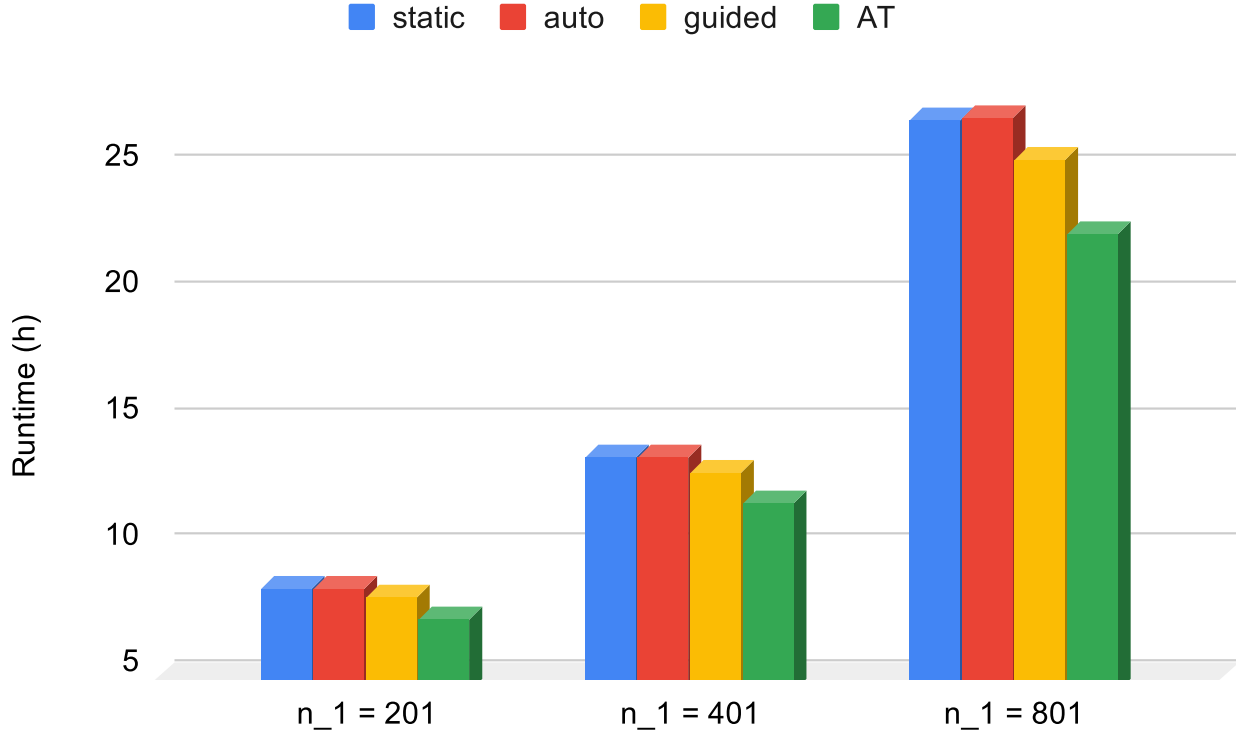}
    		\caption{16-shot RTM run time for the proposed auto-tuning, compared with the auto and the static scheduling types, for three input sizes, ($n_1 \times n_2 \times n_3$) equal to $201\times401\times401$, $401\times401\times401$, and $801\times401\times401$. These tests were performed at Ogun. Each point is a median of at least five executions.}
    		\label{fig:WorkloadComparison}
		\end{figure}

        Fig. \ref{fig:WorkloadComparison} details the results shown in Table \ref{tab:workloadComparison} for a 16-shot RTM. The proposed method outperformed the three OpenMP schedulers tested regardless of the dimension of the velocity model. Moreover, the employment of the proposed method leads to higher performance increase as the dimension of the velocity model increases.
        
        Regarding the number of shots, for most of the cases in Tables \ref{tab:machineComparison} and \ref{tab:workloadComparison}, the performance of the proposed method decreases as the number of shots increases. That indicates that the optimized chunk size for the first shot may not be optimal for the next ones and, thus, the proposed auto-tuning method could be applied to each shot in order to increase its efficiency. However, this topic is a matter of further investigation.
    
    \subsubsection{Cache misses analysis}
        The following set of performance analysis experiments aimed to answer why optimizing the chunk size through the proposed method increases the parallel efficiency of a single shot RTM. For that, we employed the HPCToolkit performance tools \cite{Adhianto2010} to measure the amount of cache misses for the three levels of cache available in NPAD, namely, L1, L2, and L3.
        
        \begin{figure}
    		\centering
		    \includegraphics[width=0.99\columnwidth]{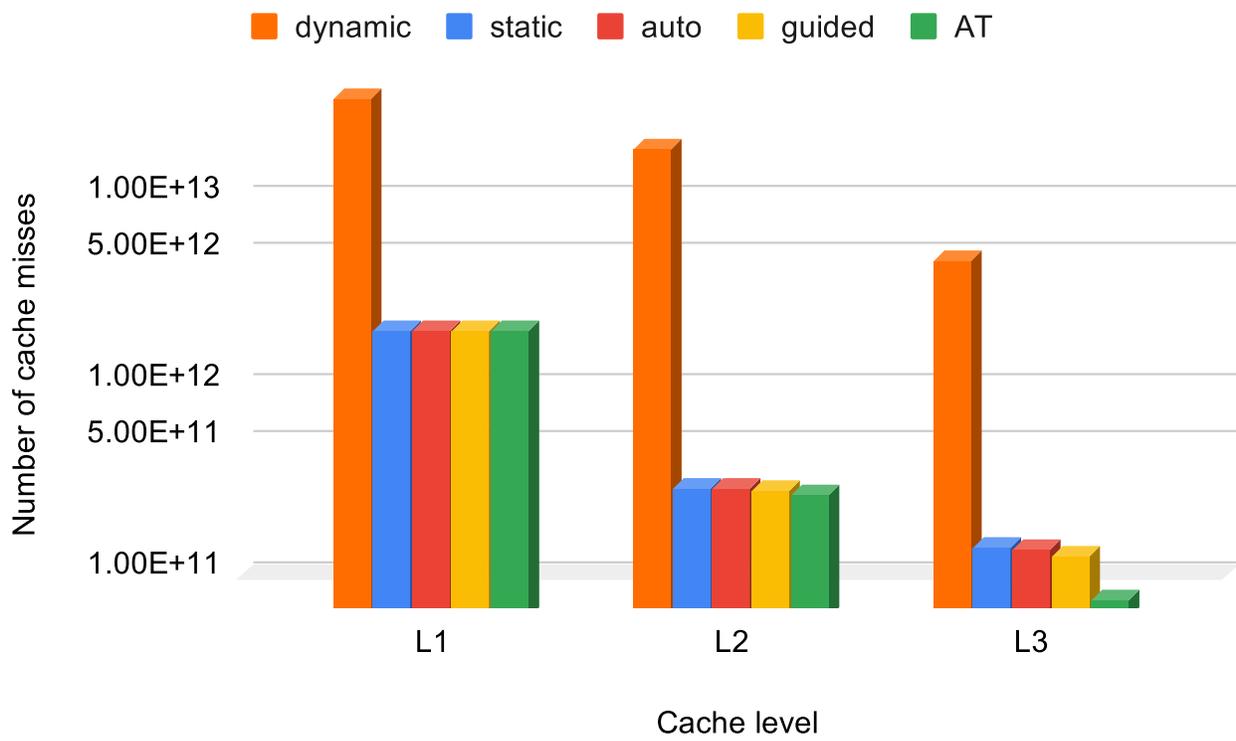}
		    \caption{Absolute number of cache misses measured by HPCToolkit for the execution of a single shot RTM using the proposed auto-tuning (AT) and the OpenMP schedulers, namely, \texttt{auto}, \texttt{static}, \texttt{dynamic}, and \texttt{guided}. For the OpenMP schedulers, the chunk size was not explicitly specified. The size of the velocity model was $(n_1\times n_2\times n_3)=401\times401\times401$. These measurements were taken at NPAD.}
		    \label{fig:cacheMisses}
		\end{figure}
        
        Fig. \ref{fig:cacheMisses} shows that using the OpenMP \texttt{dynamic} scheduler without specifying the chunk size leads to a significantly higher number of cache misses when compared to the other schedulers. This result, as well as the experiments presented by \cite{Souza-de-Assis2014b,Fernandes2018,Barros2018}, shows that using the OpenMP \texttt{dynamic} scheduler with the default unitary chunk size and other very small chunk sizes for the RTM leads to loss of performance. The reasons for that are the high number of cache misses due to false sharing, and the overhead to manage the distribution of tasks. Given that, we decided to restrict the chunk size search domain of our method to start at $50$, as mentioned before. Also, for the same reason, we omitted the results of the \texttt{dynamic} scheduler with the default chunk size from the other tests.
        
        The number of cache misses for the \texttt{auto}, \texttt{static}, and \texttt{guided} OpenMP schedulers with the default chunk size were nearly the same for the experiment shown in Fig. \ref{fig:cacheMisses} considering cache levels L1 and L2. For cache level L3, the \texttt{guided} scheduler has $7.46\%$ and $8.15\%$ less cache misses compared to \texttt{auto} and \texttt{static} schedulers respectively. The proposed method has $1\%$, $7\%$, and $43\%$ less cache misses compared to the \texttt{guided} scheduler for cache levels L1, L2, and L3 respectively.
        
    \subsubsection{Overhead analysis}
        The following set of performance analysis experiments aimed to measure the overhead of the proposed auto-tuning in an RTM. For that, we employed the MPI function \texttt{MPI\_Wtime} before and after Line \ref{l:atcall} of Algorithm \ref{alg:rtm}. In these experiments, we verified how the proposed method's overhead behaves when varying the size of the problem. For an RTM, the size of the problem can vary by either changing the size of the velocity model or the number of shots.
        
        Table \ref{tab:overhead-sizes} presents the overhead of the proposed method when varying the size of the velocity model in two computer environments, namely, NPAD and Ogun. For all experiments of this set, the overhead was inferior to $2\%$. Table \ref{tab:overhead-sizes} shows that, for this experiment, the overhead does not change significantly when the size of the velocity model changes. As the proposed method performs its iterations using the same size of the model used by the RTM iterations, the size of the model should not affect the technique's overhead.
        
        \begin{table}
            \caption{Proposed auto-tuning overhead running over a single shot of the RTM in two different computer environments (NPAD, and Ogun), and three input sizes, ($n_1 \times n_2 \times n_3$) equal to $201\times401\times401$, $401\times401\times401$, and $801\times401\times401$. Each value is the median of at least five executions.}
            \centering
            \begin{tabular}{|c|c|c|c|}
                \hline
                ~ & $\mathbf{n_1 = 201}$ & $\mathbf{n_1 = 401}$ & $\mathbf{n_1 = 801}$ \\ \hline
                \textbf{NPAD} & 1.33\% & 1.24\% & 1.28\% \\ \hline
                \textbf{Ogun} & 0.78\% & 0.84\% & 0.86\% \\ \hline
            \end{tabular}
            \label{tab:overhead-sizes}
        \end{table}
        
        Table \ref{tab:overhead-machines} shows the overhead of the proposed method when varying the number of shots in five computer environments, namely, Leuven, Yemoja, SDumont, NPAD, and Ogun. Again, for all experiments of this set, the overhead was inferior to $2\%$. The proposed method's overhead decreases as the number of shots increases because the auto-tuning is executed only for each node's first shot. The optimized chunk size is employed for all shots processed in the same node.
        
        \begin{table}
            \caption{Proposed auto-tuning overhead running over an RTM for five amounts of shots ($1$, $2$, $4$, $8$, and $16$) and five computer environments (Leuven, Yemoja, SDumont, NPAD, and Ogun). The size of the velocity model employed was $(n_1\times n_2\times n_3) = (401\times401\times401)$. Each cell is the median of at least five executions.}
            \setlength{\tabcolsep}{4.5pt} % Default value: 6pt
            \centering
            \begin{tabular}{|c|c|c|c|c|c|c|}
                \cline{3-7}
                \multicolumn{2}{c|}{\multirow{2}{*}{}} & \multicolumn{5}{c|}{\textbf{Number of shots}} \\ \cline{3-7}
                \multicolumn{2}{c|}{} & $\mathbf{1}$ & $\mathbf{2}$ & $\mathbf{4}$ & $\mathbf{8}$ & $\mathbf{16}$ \\ \hline
                \multirow{5}{*}{\begin{tabular}{c}\setlength{\tabcolsep}{0pt}\textbf{Computer}\\\textbf{environments}\end{tabular}} & \textbf{Leuven} & $1.96\%$ & $0.99\%$ & $0.50\%$ & $0.27\%$ & $0.16\%$ \\ \cline{2-7}
                & \textbf{Yemoja} & $0.76\%$ & $0.38\%$ & $0.20\%$ & $0.10\%$ & $0.05\%$ \\ \cline{2-7}
                & \textbf{SDumont} & $0.72\%$ & $0.40\%$ & $0.21\%$ & $0.11\%$ & $0.06\%$ \\ \cline{2-7}
                & \textbf{NPAD} & $1.24\%$ & $0.70\%$ & $0.37\%$ & $0.19\%$ & $0.10\%$ \\ \cline{2-7}
                & \textbf{Ogun} & $0.84\%$ & $0.46\%$ & $0.24\%$ & $0.12\%$ & $0.06\%$ \\ \hline
            \end{tabular}
            \label{tab:overhead-machines}
        \end{table}

\section{Related work}
\label{sec:literature}
    % AT but not RUN-TIME:
    Several works have introduced auto-tuning techniques on multicore systems. Katagiri \textit{et al}. \cite{Katagiri2014,Katagiri2015} presented ppOpen-AT, a framework for code optimization guided by directives. Sena \textit{et al}. \cite{Sena2011} proposed a method to determine a near-optimal workload chunk size by testing a set of possible values in a few time steps of an RTM and choosing the one with the shortest execution time. Andreolli \textit{et al}. \cite{Andreolli2014,Andreolli2015} introduced an approach to tune seismic applications automatically by compiling and running each set of parameters chosen by a genetic algorithm, including chunk size and compilation flags. Kamil \textit{et al}. \cite{Kamil2010} presented a framework to generate auto-tuned C, Fortran, and CUDA codes for stencil applications specified in sequential Fortran 95. In common, these works \cite{Katagiri2014,Katagiri2015,Sena2011,Andreolli2014,Andreolli2015,Kamil2010} perform the auto-tuning before the execution time. By doing so, these methods may find a system status different from the tuning time as aspects such as memory availability may change over time. Conversely, the proposed method performs auto-tuning at run time, enabling it to capture more realistic parameters.
    
    % AT+RUN-TIME but not OpenMP:
    Many authors have proposed run-time auto-tuning approaches. Padoin \textit{et al}. \cite{Padoin2014,Padoin2017} employed load balancing techniques along with processors frequency control tools to improve the energy efficiency of imbalanced parallel applications in multicore systems. Both works use processors frequency scaling techniques to slow down less loaded cores to save energy. Tchiboukdjian \textit{et al}. \cite{Tchiboukdjian2011} introduced a scheduler for applications with linear access to shared memory. They aim to improve locality by guaranteeing that all data in the cache are used before being replaced. Olivier \textit{et al}. \cite{Olivier2011,Olivier2012} proposed an OpenMP hierarchical task scheduler for multicore systems using a work-stealing strategy \cite{Blumofe1999}. Different from \cite{Padoin2014,Padoin2017,Tchiboukdjian2011,Olivier2011,Olivier2012}, the proposed method focuses on tuning the scheduling of the OpenMP directive for parallel loops automatically. According to Diaz \textit{et al}. \cite{Diaz2012}, OpenMP is a predominant approach in shared memory architectures both in industry and academia, which enforces the relevance of our work.
    
    %AT+RUN-TIME+OPENMP but not META-HEURISTIC:
    Some works also introduced auto-tuning techniques which can schedule parallel loops using OpenMP. Kale and Gropp \cite{Kale2010} employed a strategy combining static and dynamic scheduling to tune implementations of regular meshes applications in clusters of symmetric multiprocessing (SMP) machines and by Donfack \textit{et al}. \cite{Donfack2012} to tune highly-optimized dense matrix factorization methods. Kale \textit{et al} \cite{Kale2014} proposed improving a combined static and dynamic scheduling on SMP machines by enforcing spatial locality. Bak \textit{et al} \cite{Bak2018} introduce a load-balancing method based on the integration of Charm++ and OpenMP to dynamically distribute user-created tasks. Differently, we employ CSA to automatically determine the size of the chunks of parallel loop iterations. The relation between chunk size and total execution time in geophysical methods is unknown but has been shown to have many local optima \cite{Fernandes2018,Barros2018}. By using a global optimization strategy, our method aims to increase the probability of finding the global optimum.
    
    %AT+RUN-TIME+OPENMP+META-HEURISTIC but not RTM:
    Tiwari and Hollingsworth \cite{Tiwari2011} proposed an online auto-tuning method to tune programmer-defined parameters. For that, a set of nodes of a distributed system generate, compile, and run code versions using different sets of parameters. The optimization of the sets of parameters is defined via simplex. Differently from our work, Tiwari and Hollingsworth's method relies on a distributed system. Although it can be performed in a single node, the results presented by Tiwari and Hollingsworth show that the simplex did not converge and increased the time of execution when employing a single node. Also, the overhead of Tiwari and Hollingsworth's method increases as the size of the problem decreases while our results show that our method's overhead is not correlated to the size of the problem.
    
    We have shown before that CSA \cite{xavier2010coupled} is a promising method to obtain a near-optimal load balance for a 3D FDM. Experiments in \cite{Fernandes2018,Barros2018} showed that the optimal load balancing for shared memory environments depends on the hardware and software employed. In this paper, we use the auto-tuning method proposed in our previous works \cite{Fernandes2018,Barros2018} to the RTM and provide overhead, parallel performance, and CSA parameters analysis.

\section{Conclusions}
\label{sec:conclusions}
    We have proposed a CSA-based auto-tuning strategy for seeking the optimal chunk size of OpenMP's dynamic loop scheduler to reduce the execution time of a 3D reverse time migration algorithm. The proposed approach is designed to work wherever the code is executed, being robust for changes in computational environment parameters, such as the number of threads, processors, memory hierarchy, and compiler.
    
    We showed that the proposed auto-tuning reduces the number of cache misses compared to the default OpenMP schedulers. Also, for all experiments performed in this paper, the overhead of the proposed method was inferior to $2\%$.
    
    Experiments running an RTM showed that, in most cases, the proposed auto-tuning method outperforms the OpenMP \texttt{static}, \texttt{auto}, and \texttt{guided} schedulers in different computational resources for different amounts of seismic shots. For this set of experiments, the proposed auto-tuning method reached speedups up to $23.1\%$.
    
    Another set of tests with RTM presented the proposed method's performance in comparison with the OpenMP \texttt{static}, \texttt{auto}, and \texttt{guided} schedulers for a varying input size and a different number of seismic shots. For this set of experiments, the proposed auto-tuning outperformed the OpenMP schedulers, reaching up to $33\%$ speedup. The proposed method performed better for larger inputs, which improves the parallel scalability of 3D RTM.
    
    In summary, the proposed method has shown to be robust and scalable for the 3D RTM. As future work, we are interested in testing the proposed method's ability to improve the performance of similar wave-based algorithms, such as full-waveform inversion (FWI) and other iterative applications.
    
\section{ACKNOWLEDGMENTS}
    The authors gratefully acknowledge support from Shell Brazil through the project ``\textit{Novos M\'etodos de Explora\c{c}\~ao S\'ismica por Invers\~ao Completa das Formas de Onda}'' at the Universidade Federal do Rio Grande do Norte (UFRN) and the strategic importance of the support given by ANP through the R\&D levy regulation.
    The authors also acknowledge the National Laboratory for Scientific Computing (LNCC/MCTI, Brazil), the High-Performance Computing Center at UFRN (NPAD/UFRN), the Center of Technology of the UFRN (CT/UFRN), and the Manufacturing and Technology Integrated Campus of the National Service of Industrial Training (SENAI CIMATEC) for providing HPC resources of the SDumont, NPAD, Leuven, Yemoja, and Ogun supercomputers, which have contributed to the research results reported within this paper.
    Finally, the authors thank Jorge Lopez from Shell and the anonymous reviewers for providing essential comments on this paper.

\bibliographystyle{IEEEtran}
\bibliography{at}

\begin{thebibliography}{10}
\providecommand{\url}[1]{#1}
\csname url@samestyle\endcsname
\providecommand{\newblock}{\relax}
\providecommand{\bibinfo}[2]{#2}
\providecommand{\BIBentrySTDinterwordspacing}{\spaceskip=0pt\relax}
\providecommand{\BIBentryALTinterwordstretchfactor}{4}
\providecommand{\BIBentryALTinterwordspacing}{\spaceskip=\fontdimen2\font plus
\BIBentryALTinterwordstretchfactor\fontdimen3\font minus
  \fontdimen4\font\relax}
\providecommand{\BIBforeignlanguage}[2]{{%
\expandafter\ifx\csname l@#1\endcsname\relax
\typeout{** WARNING: IEEEtran.bst: No hyphenation pattern has been}%
\typeout{** loaded for the language `#1'. Using the pattern for}%
\typeout{** the default language instead.}%
\else
\language=\csname l@#1\endcsname
\fi
#2}}
\providecommand{\BIBdecl}{\relax}
\BIBdecl

\bibitem{Kearey2002}
P.~Kearey, M.~Brooks, and I.~Hill, \emph{{An introduction to geophysical
  exploration}}, 3rd~ed.\hskip 1em plus 0.5em minus 0.4em\relax Malden, MA :
  Blackwell Science, 2002.

\bibitem{Baysal1983}
E.~Baysal, D.~D. Kosloff, and J.~W.~C. Sherwood, ``{Reverse time migration},''
  \emph{Geophysics}, vol.~48, no.~11, pp. 1514--1524, 1983.

\bibitem{Kosloff1983}
D.~D. Kosloff and E.~Baysal, ``{Migration with the full acoustic wave
  equation},'' \emph{Geophysics}, vol.~48, no.~6, pp. 677--687, 1983.

\bibitem{Zhang2009}
{Zhang, Jin-Hai and Wang, Shu-Qin and Yao, Zhen-Xing}, ``{Accelerating 3D
  Fourier migration with graphics processing units},'' \emph{GEOPHYSICS}, 2009.

\bibitem{Araya-Polo2009}
M.~Araya-Polo, F.~Rubio, R.~{De La Cruz}, M.~Hanzich, J.~M. Cela, and D.~P.
  Scarpazza, ``{3D seismic imaging through reverse-time migration on
  homogeneous and heterogeneous multi-core processors},'' \emph{Scientific
  Programming}, 2009.

\bibitem{Nunes-do-rosario2015}
D.~A. Nunes-Do-Ros{\'{a}}rio, S.~Xavier-De-Souza, R.~C. Maciel, and J.~C.
  Costa, ``{Parallel Scalability of a Fine-Grain Prestack Reverse Time
  Migration Algorithm},'' \emph{IEEE Geoscience and Remote Sensing Letters},
  2015.

\bibitem{naono2010software}
K.~Naono, K.~Teranishi, J.~Cavazos, and R.~Suda, \emph{Software automatic
  tuning: from concepts to state-of-the-art results}.\hskip 1em plus 0.5em
  minus 0.4em\relax Springer Science \& Business Media, 2010.

\bibitem{Demetrios2020}
\BIBentryALTinterwordspacing
A.~M. Coutinho~Demetrios, D.~De~Sensi, A.~F. Lorenzon, K.~Georgiou,
  J.~Nunez-Yanez, K.~Eder, and S.~Xavier-de Souza, ``Performance and energy
  trade-offs for parallel applications on heterogeneous multi-processing
  systems,'' \emph{Energies}, vol.~13, no.~9, p. 2409, May 2020. [Online].
  Available: \url{http://dx.doi.org/10.3390/en13092409}
\BIBentrySTDinterwordspacing

\bibitem{OpemMP}
L.~{Dagum} and R.~{Menon}, ``Open{MP}: an industry standard {API} for
  shared-memory programming,'' \emph{IEEE Computational Science and
  Engineering}, vol.~5, no.~1, pp. 46--55, Jan 1998.

\bibitem{xavier2010coupled}
{Xavier-de-Souza, Samuel and Suykens, Johan AK and Vandewalle, Joos and
  Boll{\'e}, D{\'e}sir{\'e}}, ``Coupled simulated annealing,'' \emph{IEEE
  Transactions on Systems, Man, and Cybernetics, Part B (Cybernetics)},
  vol.~40, no.~2, pp. 320--335, 2010.

\bibitem{Carcione2002}
J.~M. Carcione, G.~C. Herman, and A.~P.~E. ten Kroode, ``{Seismic modeling},''
  \emph{Geophysics}, 2002.

\bibitem{Cerjan1985}
C.~Cerjan, D.~Kosloff, R.~Kosloff, and M.~Reshef, ``{A nonreflecting boundary
  condition for discrete acoustic and elastic wave equations},'' p. 705, 1985.

\bibitem{yilmaz2001b}
{\"{O}}.~Yilmaz, \emph{{Seismic Data Analysis: Processing, Inversion, and
  Interpretation of Seismic Data}}, ser. Investigations in geophysics.\hskip
  1em plus 0.5em minus 0.4em\relax Society of Exploration Geophysicists, 2001,
  no. v. 2.

\bibitem{MPI1994}
L.~Clarke, I.~Glendinning, and R.~Hempel, ``The {MPI} message passing interface
  standard,'' in \emph{Programming Environments for Massively Parallel
  Distributed Systems}, K.~M. Decker and R.~M. Rehmann, Eds.\hskip 1em plus
  0.5em minus 0.4em\relax Basel: Birkh{\"a}user Basel, 1994, pp. 213--218.

\bibitem{kirkpatrick1983optimization}
S.~Kirkpatrick, C.~D. Gelatt, and M.~P. Vecchi, ``Optimization by simulated
  annealing,'' \emph{Science}, vol. 220, no. 4598, pp. 671--680, 1983.

\bibitem{gonccalves2018parallel}
{Gon{\c{c}}alves-e-Silva, Kayo and Aloise, Daniel and Xavier-de-Souza, Samuel},
  ``Parallel synchronous and asynchronous coupled simulated annealing,''
  \emph{The Journal of Supercomputing}, vol.~74, no.~6, pp. 2841--2869, 2018.

\bibitem{Assis2019}
\BIBentryALTinterwordspacing
{Assis, Italo A S and Oliveira, Antonio D S and Barros, Tiago and Sardina,
  Idalmis M and Bianchini, Calebe P and Xavier-de-Souza, Samuel},
  ``{Distributed-Memory Load Balancing With Cyclic Token-Based Work-Stealing
  Applied to Reverse Time Migration},'' \emph{IEEE Access}, vol.~7, pp.
  128\,419--128\,430, 2019. [Online]. Available:
  \url{https://ieeexplore.ieee.org/document/8822671/}
\BIBentrySTDinterwordspacing

\bibitem{wang2015frequencies}
Y.~Wang, ``Frequencies of the {R}icker wavelet,'' \emph{Geophysics}, vol.~80,
  no.~2, pp. A31--A37, 2015.

\bibitem{Symes2007}
W.~Symes, ``{Reverse time migration with optimal checkpointing},''
  \emph{Geophysics}, 2007.

\bibitem{Griewank2000}
A.~Griewank and A.~Walther, ``{Algorithm 799: revolve: an implementation of
  checkpointing for the reverse or adjoint mode of computational
  differentiation},'' \emph{ACM Transactions on Mathematical Software}, 2000.

\bibitem{fiber}
T.~Katagiri, K.~Kise, H.~Honda, and T.~Yuba, ``Fiber: A generalized framework
  for auto-tuning software,'' in \emph{High Performance Computing},
  A.~Veidenbaum, K.~Joe, H.~Amano, and H.~Aiso, Eds.\hskip 1em plus 0.5em minus
  0.4em\relax Berlin, Heidelberg: Springer Berlin Heidelberg, 2003, pp.
  146--159.

\bibitem{Furtunato2020}
A.~F.~A. {Furtunato}, K.~{Georgiou}, K.~{Eder}, and S.~{Xavier-de-Souza},
  ``When parallel speedups hit the memory wall,'' \emph{IEEE Access}, vol.~8,
  pp. 79\,225--79\,238, 2020.

\bibitem{Barros2018}
\BIBentryALTinterwordspacing
{Barros, T. and Fernandes, J. B. and Souza-de-Assis, I. A. and Xavier-de-Souza,
  S.}, ``{Auto-Tuning of 3D Acoustic Wave Propagation in Shared Memory
  Environments},'' in \emph{First EAGE Workshop on High Performance Computing
  for Upstream in Latin America}.\hskip 1em plus 0.5em minus 0.4em\relax
  Santander: EarthDoc, 2018. [Online]. Available:
  \url{http://www.earthdoc.org/publication/publicationdetails/?publication=94579}
\BIBentrySTDinterwordspacing

\bibitem{Openmp2015}
\BIBentryALTinterwordspacing
{OpenMP Architecture Review Board}, ``{{\{}OpenMP{\}} Application Program
  Interface Version 4.5},'' 2015. [Online]. Available:
  \url{https://www.openmp.org/wp-content/uploads/openmp-4.5.pdf}
\BIBentrySTDinterwordspacing

\bibitem{Libgomp2019}
\BIBentryALTinterwordspacing
{GCC Team}. (2019, jun) {GNU libgomp}. [Online]. Available:
  \url{https://gcc.gnu.org/onlinedocs/libgomp/}
\BIBentrySTDinterwordspacing

\bibitem{Gitlibgomp2020}
\BIBentryALTinterwordspacing
------. (2020, jun) {GNU Compiler Collection (GCC)}. [Online]. Available:
  \url{https://github.com/gcc-mirror/gcc/blob/master/libgomp/loop.c}
\BIBentrySTDinterwordspacing

\bibitem{de1960modification}
A.~De~Hoop, ``A modification of {C}agniard's method for solving seismic pulse
  problems,'' \emph{Applied Scientific Research, Section B}, vol.~8, no.~1, pp.
  349--356, 1960.

\bibitem{Adhianto2010}
L.~Adhianto, S.~Banerjee, M.~Fagan, M.~Krentel, G.~Marin, J.~Mellor-Crummey,
  and N.~R. Tallent, ``{HPCTOOLKIT: Tools for performance analysis of optimized
  parallel programs},'' \emph{Concurrency Computation Practice and Experience},
  2010.

\bibitem{Souza-de-Assis2014b}
{Souza-de-Assis, Italo A. and Nunes-do-Ros{\'{a}}rio, D. A. and Maciel, R. C.
  and Xavier-de-Souza, S.}, ``{Um Algoritmo Paralelo Eficiente de
  Propaga{\c{c}}{\~{a}}o de Onda Ac{\'{u}}stica 3D},'' in \emph{Rio Oil {\&}
  Gas Expo and Conference}, Rio de Janeiro, 2014.

\bibitem{Fernandes2018}
\BIBentryALTinterwordspacing
{Fernandes, Jo{\~{a}}o B. and Souza-de-Assis, {\'{I}}talo A. and Barros, Tiago
  and Xavier-de-Souza, Samuel}, ``{Automatic Scheduler for 3D Seismic Modeling
  by Finite Differences},'' in \emph{Rio Oil {\&} Gas}, Rio de Janeiro, 2018.
  [Online]. Available:
  \url{https://stt.ibp.org.br/eventos/2018/riooil2018/pdfs/Riooil2018\_1901\_201806151345riooeg\_end\_paper.pdf}
\BIBentrySTDinterwordspacing

\bibitem{Katagiri2014}
T.~Katagiri, S.~Ohshima, and M.~Matsumoto, ``{Auto-tuning of computation
  kernels from an FDM code with ppOpen-AT},'' in \emph{Proceedings - 2014 IEEE
  8th International Symposium on Embedded Multicore/Manycore SoCs, MCSoC 2014},
  2014.

\bibitem{Katagiri2015}
------, ``{Directive-Based Auto-Tuning for the Finite Difference Method on the
  Xeon Phi},'' in \emph{Proceedings - 2015 IEEE 29th International Parallel and
  Distributed Processing Symposium Workshops, IPDPSW 2015}, 2015.

\bibitem{Sena2011}
A.~C. Sena, A.~P. Nascimento, C.~Boeres, V.~Rebello, and A.~Bulc{\~{a}}o, ``{An
  approach to optimise the execution of RTM algorithm in multicore machines},''
  in \emph{Proceedings - 2011 7th IEEE International Conference on eScience,
  eScience 2011}, 2011.

\bibitem{Andreolli2014}
C.~Andreolli, P.~Thierry, L.~Borges, C.~Yount, and G.~Skinner, ``{Genetic
  Algorithm Based Auto-Tuning of Seismic Applications on Multi and Manycore
  Computers},'' in \emph{EAGE Workshop on High Performance Computing for
  Upstream}, 2014.

\bibitem{Andreolli2015}
\BIBentryALTinterwordspacing
C.~Andreolli, P.~Thierry, L.~Borges, G.~Skinner, and C.~Yount,
  ``{Characterization and Optimization Methodology Applied to Stencil
  Computations},'' in \emph{High Performance Parallelism Pearls}, J.~Jeffers
  and J.~Reinders, Eds.\hskip 1em plus 0.5em minus 0.4em\relax Boston:
  Elsevier, 2015, ch.~23, pp. 377--396. [Online]. Available:
  \url{http://www.sciencedirect.com/science/article/pii/B9780128021187000236}
\BIBentrySTDinterwordspacing

\bibitem{Kamil2010}
S.~{Kamil}, C.~{Chan}, L.~{Oliker}, J.~{Shalf}, and S.~{Williams}, ``An
  auto-tuning framework for parallel multicore stencil computations,'' in
  \emph{2010 IEEE International Symposium on Parallel Distributed Processing
  (IPDPS)}, 2010, pp. 1--12.

\bibitem{Padoin2014}
E.~L. Padoin, M.~Castro, L.~L. Pilla, P.~O.~A. Navaux, and J.~M{\'{e}}haut,
  ``{Saving energy by exploiting residual imbalances on iterative
  applications},'' in \emph{2014 21st International Conference on High
  Performance Computing (HiPC)}, 2014, pp. 1--10.

\bibitem{Padoin2017}
E.~L. Padoin, L.~L. Pilla, M.~Castro, P.~O. Navaux, and J.~F. M{\'{e}}haut,
  ``{Exploration of load balancing thresholds to save energy on iterative
  applications},'' in \emph{Communications in Computer and Information
  Science}, 2017.

\bibitem{Tchiboukdjian2011}
M.~Tchiboukdjian, V.~Danjean, T.~Gautier, F.~{Le Mentec}, and B.~Raffin, ``{A
  work stealing scheduler for parallel loops on shared cache multicores},'' in
  \emph{Lecture Notes in Computer Science (including subseries Lecture Notes in
  Artificial Intelligence and Lecture Notes in Bioinformatics)}, 2011.

\bibitem{Olivier2011}
S.~L. Olivier, A.~K. Porterfield, K.~B. Wheeler, and J.~F. Prins, ``{Scheduling
  task parallelism on multi-socket multicore systems},'' in \emph{Proceedings
  of the 1st International Workshop on Runtime and Operating Systems for
  Supercomputers, ROSS 2011}, 2011.

\bibitem{Olivier2012}
\BIBentryALTinterwordspacing
S.~L. Olivier, A.~K. Porterfield, K.~B. Wheeler, M.~Spiegel, and J.~F. Prins,
  ``Open{MP} task scheduling strategies for multicore {NUMA} systems,''
  \emph{The International Journal of High Performance Computing Applications},
  vol.~26, no.~2, pp. 110--124, 2012. [Online]. Available:
  \url{https://doi.org/10.1177/1094342011434065}
\BIBentrySTDinterwordspacing

\bibitem{Blumofe1999}
\BIBentryALTinterwordspacing
R.~D. Blumofe and C.~E. Leiserson, ``Scheduling multithreaded computations by
  work stealing,'' \emph{J. ACM}, vol.~46, no.~5, pp. 720--748, Sep. 1999.
  [Online]. Available: \url{http://doi.acm.org/10.1145/324133.324234}
\BIBentrySTDinterwordspacing

\bibitem{Diaz2012}
J.~Diaz, C.~Mu{\~{n}}oz-Caro, and A.~Ni{\~{n}}o, ``{A survey of parallel
  programming models and tools in the multi and many-core era},'' in \emph{IEEE
  Transactions on Parallel and Distributed Systems}, 2012.

\bibitem{Kale2010}
V.~Kale and W.~Gropp, ``Load balancing for regular meshes on {SMP}s with
  {MPI},'' in \emph{Recent Advances in the Message Passing Interface},
  R.~Keller, E.~Gabriel, M.~Resch, and J.~Dongarra, Eds.\hskip 1em plus 0.5em
  minus 0.4em\relax Berlin, Heidelberg: Springer Berlin Heidelberg, 2010, pp.
  229--238.

\bibitem{Donfack2012}
S.~{Donfack}, L.~{Grigori}, W.~D. {Gropp}, and V.~{Kale}, ``Hybrid
  static/dynamic scheduling for already optimized dense matrix factorization,''
  in \emph{2012 IEEE 26th International Parallel and Distributed Processing
  Symposium}, 2012, pp. 496--507.

\bibitem{Kale2014}
V.~Kale, A.~Randles, and W.~D. Gropp, ``Locality-optimized mixed static/dynamic
  scheduling for improving load balancing on {SMP}s,'' in \emph{Proceedings of
  the 21st European MPI Users' Group Meeting}, 2014, pp. 115--116.

\bibitem{Bak2018}
S.~{Bak}, H.~{Menon}, S.~{White}, M.~{Diener}, and L.~{Kale}, ``Multi-level
  load balancing with an integrated runtime approach,'' in \emph{2018 18th
  IEEE/ACM International Symposium on Cluster, Cloud and Grid Computing
  (CCGRID)}, 2018, pp. 31--40.

\bibitem{Tiwari2011}
A.~{Tiwari} and J.~K. {Hollingsworth}, ``Online adaptive code generation and
  tuning,'' in \emph{2011 IEEE International Parallel Distributed Processing
  Symposium}, 2011, pp. 879--892.

\end{thebibliography}

%\EOD

\end{document}